\documentclass[aps,graphicx,twocolumn,showpacs]{revtex4}
\usepackage{amscd}
\usepackage{graphicx}

\begin{document}

\title{Efficient faithful qubit transmission with frequency degree of freedom\footnote{Published in
Opt. Commun. 282 (2009) 4025-4027}}
\author{ Xi-Han Li$^{a}$, Bao-Kui Zhao$^{a}$, Yu-Bo Sheng$^{a}$,
 Fu-Guo Deng$^{a,b}$\footnote {Author to whom correspondence should be
 addressed.\\
Email address: fgdeng@bnu.edu.cn},  Hong-Yu Zhou$^{a}$}
\address{$^a$Key Laboratory of Beam Technology and Material
Modification of Ministry of Education, and College of Nuclear
Science and Technology, Beijing Normal University,
Beijing 100875,  China\\
$^b$Department of Physics,  Beijing Normal University, Beijing
100875, China }
\date{\today}

\begin{abstract}
We propose an efficient faithful polarization-state transmission
scheme by utilizing frequency degree of freedom besides polarization
and an additional qubit prepared in a fixed polarization. An
arbitrary single-photon polarization state is protected against the
collective noise probabilistically. With the help of frequency beam
splitter  and frequency shifter, the success probability of our
faithful qubit transmission scheme with frequency degree of freedom
can be 1/2 in principle.
\end{abstract}
 \maketitle

\section{introduction}

The main task of quantum communication is transmitting information
between two distant parties. However, during the transmission, the
quantum information carriers are usually infected by various things,
which we call noise. By far, various methods have been proposed to
solve the problem caused by noise. For example,  phase coding was
introduced \cite{phase} to overcome the influence on the
polarization of photons by thermal fluctuation, vibration and the
imperfection of the fiber. Corresponding to different types of
noise, some other methods are proposed  such as quantum error
correct code (QEC) \cite{qec}, error rejection
\cite{pla343,apl91,prl95} and decoherence-free subspace (DFS)
\cite{dfs1,dfs2,dfs3,dfs4}. Although QEC and DFS methods can be used
to suppress  the effect of noise effectively, they are sensitive to
channel losses and need much resource.

As the noise usually fluctuates slowly in time, there is an
important precondition for dealing with it in quantum information
processing, which called a collective noise assumption
\cite{collective}. That is, if several photons travel through the
collective-noise channel simultaneously or the maximum interval
between them is shorter than the variation of noise, the alteration
caused by noise is the same one for each qubit. Recently, some
error-rejection qubit transmission schemes have been proposed by
using only one or two photons with linear optics. In 2005, Kalamidas
proposed two single-photon transmission schemes to reject arbitrary
errors \cite{pla343}. The average success probability of the second
protocol is 100\%. However, fast polarization modulator (Pockels
cell), whose synchronization makes it difficult to be implemented
with current technology, are employed. Subsequently, we presented a
single-photon transmission scheme against collective noise with only
passive linear optical elements \cite{apl91}. In 2005, Yamamoto et
al proposed a qubit distribution scheme against collective noise
with the help of one additional qubit in a fixed polarization
\cite{prl95}. We call it PRL95 protocol below, whose experimental
results with a collective dephasing noise are published in
Ref.\cite{njp}. The total success probability can be improved from
1/16 to 1/8 with "deterministic two-qubit operations", which was not
presented clearly in their work \cite{prl95}. At present, a
deterministic two-qubit operation based on linear optics is
difficult to be implemented in practice as it can be done with only
a very low efficiency \cite{cnot}.

In recent years, some other degrees of freedom (DOFs) of photons
besides their polarization attract much attention, such as the
frequency \cite{fre}, spatial mode \cite{simon,shengpra}, orbital
angular momentum (OAM)\cite{oam}, transverse spatial mode
\cite{transverse} of photons, and so on. Additional to polarization,
other DOFs can be used to implement complete Bell-state analysis
\cite{bell}, entanglement purification \cite{simon,shengpra,fre},
superdense coding \cite{dense}, and so on. As the frequency DOF is
stable in any transmission surroundings, it was used to code message
in quantum key distribution schemes and some good results were
obtained \cite{freqkd1,freqkd2,freqkd3,freqkd4}. Photon with more
than one degree of freedom (DOF) can be prepared with current
technology easily \cite{prepare}.

Quantum state transmission  is important to quantum information
process as the transmission of qubits carrying information is the
first step of quantum communication. In this paper, we proposed an
efficient faithful qubit transmission scheme assisted by an
additional qubit which has different frequency with the signal one.
The frequency DOF is used to mark these two photons. The success
probability of our protocol is eight times of PRL95 protocol at best
without using the deterministic two-qubit operation. We also find
that the success probability of PRL95 scheme can be improved from
1/16 to 1/8 just by adding a half-wave plate (HWP).

\section{Efficient faithful quantum state transmission with frequency degree of freedom}

The principle of our scheme for efficient faithful quantum state
transmission with frequency degree of freedom is shown in Fig.1. An
arbitrary single-photon pure state to be transmitted can be written
as
\begin{eqnarray}
\vert \psi \rangle_s=\alpha \vert H \rangle_s + \beta \vert V
\rangle_s, \,\,\,\,\,\,(|\alpha|^2+ |\beta|^2=1).
\end{eqnarray}
Here $\vert H \rangle$ and $\vert V \rangle$ represent the
horizontal and the vertical polarization states, respectively. The
subscript $s$ represents the signal photon. Alice prepares a
reference photon in the state $(1/\sqrt{2})(\vert H \rangle_r +
\vert V \rangle_r)$. The frequencies of the reference photon and the
signal one are $\omega_r$ and $\omega_s$, respectively. Two photons
with different frequencies can be prepared with current technology
\cite{freqkd1,freqkd4,prepare}. The input state of the quantum
system composed of the two photons can be written as
\begin{eqnarray}
\vert \psi \rangle_{rs}= \frac{1}{\sqrt{2}}(\vert H_r, \omega_r
\rangle + \vert V_r,\omega_r \rangle) \otimes (\alpha \vert H_s,
\omega_s\rangle + \beta\vert V_s,\omega_s \rangle).\nonumber
\end{eqnarray}
For the sake of simplicity, $\omega_r$ and $\omega_s$ can be omitted
in the following, i.e.,
\begin{eqnarray}
\vert \psi \rangle_{rs}= \frac{1}{\sqrt{2}}(\vert H_r \rangle +
\vert V_r\rangle) \otimes (\alpha \vert H_s\rangle + \beta\vert
V_s\rangle).
\end{eqnarray}
The subscript $r$ and $s$ can mark the frequency difference between
the two photons. Each photon is split into two pulses by the first
polarizing beam splitter (PBS$_1$), which drives $\vert H \rangle$
going through channel 1 and $\vert V \rangle $ through channel 2.
Notice that the two photons are sent to Bob simultaneously in this
protocol. These two noise channels have equal length and the phase
shift of these two channels is consistent. The alternation caused by
noise is
\begin{eqnarray}
\vert H \rangle_{r,s} &\rightarrow &  \delta_1 \vert H \rangle_{r,s}
+ \eta_1 \vert V \rangle_{r,s},\nonumber\\
 \vert V \rangle_{r,s}
&\rightarrow & \delta_2 \vert H \rangle_{r,s} + \eta_2 \vert V
\rangle_{r,s}. \label{noise}
\end{eqnarray}
Parts of each photon converge at PBS$_2$.

\begin{figure}[!h]
\begin{center}
\includegraphics[width=8cm,angle=0]{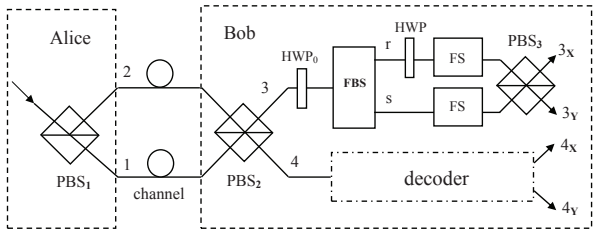}
\caption{Schematics of qubit distribution protocol with frequency
degree of freedom. Two photons with different frequencies $\omega_r$
and $\omega_s$ are sent to Bob simultaneously. The noise channels 1
and 2 have the same length. Frequency beam splitter (FBS) is used to
divide photons with different wavelengthes into different spatial
modes, i.e. the reference photon with the frequency $\omega_r$ goes
to the up path and the signal one  with the frequency $\omega_s$
goes to the down path. Frequency shifter (FS) is used to eliminate
the frequency distinguishability. The decoder set in port 4 is same
as the one in port 3, which is indicated by the dash-dotted
rectangle.}
\end{center}
\end{figure}

The state of the quantum system composed of the two photons after
the PBS$_2$ becomes
\begin{eqnarray}
\vert \psi_1\rangle &=&\frac{1}{\sqrt{2}}[\alpha \delta_1^2\vert H
\rangle_{r3}\vert H \rangle_{s3}+ \beta\eta_2^2\vert V
\rangle_{r3}\vert V \rangle_{s3}\nonumber\\&&+ \delta_1
\eta_2(\alpha \vert V \rangle_{r3} \vert H
\rangle_{s3}+\beta \vert H \rangle_{r3} \vert V \rangle_{s3})\nonumber\\
&&+ \alpha\eta_1^2\vert V \rangle_{r4}\vert V \rangle_{s4}+
\beta\delta_2^2\vert H \rangle_{r4}\vert H
\rangle_{s4}\nonumber\\&&+ \eta_1 \delta_2(\alpha\vert H
\rangle_{r4} \vert V \rangle_{s4}+\beta\vert
V \rangle_{r4} \vert H \rangle_{s4})\nonumber\\
&&+ \alpha\delta_1 \eta_1(\vert H \rangle_{r3} \vert V \rangle_{s4}
+ \vert V \rangle_{r4} \vert H \rangle_{s3})\nonumber\\&&+ \delta_1
\delta_2(\alpha\vert H \rangle_{r4}\vert H \rangle_{s3}+\beta\vert
H \rangle_{r3}\vert H \rangle_{s4}) \nonumber\\
&&+ \beta \delta_2 \eta_2(\vert H \rangle_{r4} \vert V \rangle_{s3}
+ \vert V \rangle_{r3} \vert H \rangle_{s4}) \nonumber\\&&+ \eta_1
\eta_2(\alpha\vert V \rangle_{r3}\vert V \rangle_{s4}+\beta\vert V
\rangle_{r4}\vert V \rangle_{s3})]. \label{psi1}
\end{eqnarray}
Here 3 and 4 are two output ports of PBS$_2$. A HWP whose
orientation is 45$^\circ$  is placed in the path 3 behind the
PBS$_2$, which acts as the $\sigma_x$ operation
\begin{eqnarray}
\vert H \rangle_{r,s}&\rightarrow& \vert V \rangle_{r,s}, \nonumber\\
\vert V \rangle_{r,s}&\rightarrow& \vert H \rangle_{r,s}
\end{eqnarray}
The total state before the decoder becomes
\begin{eqnarray}
\vert \psi_2\rangle &=& \frac{1}{\sqrt{2}}[\alpha \delta_1^2\vert V
\rangle_{r3}\vert V \rangle_{s3} + \beta \eta_2^2\vert H
\rangle_{r3}\vert H \rangle_{s3}\nonumber\\&&+ \alpha \eta_1^2\vert
V \rangle_{r4}\vert V \rangle_{s4} + \beta \delta_2^2\vert H
\rangle_{r4}\vert H \rangle_{s4}\nonumber\\&& +\alpha
\delta_1\eta_1(\vert V \rangle_{r3}\vert V \rangle_{s4} + \vert V
\rangle_{r4}\vert V \rangle_{s3})\nonumber\\&&+\beta \delta_2
\eta_2(\vert H \rangle_{r3}\vert H \rangle_{s4}
+ \vert H \rangle_{r4}\vert H \rangle_{s3})\nonumber\\
&&+ \delta_1 \eta_2(\alpha \vert H \rangle_{r3} \vert V
\rangle_{s3}+ \beta \vert V \rangle_{r3} \vert H
\rangle_{s3})\nonumber\\&&+ \eta_1 \delta_2(\alpha \vert H
\rangle_{r4} \vert V \rangle_{s4} + \beta
\vert V \rangle_{r4} \vert H \rangle_{s4})\nonumber\\
&&+ \delta_1 \delta_2 (\alpha \vert H \rangle_{r4} \vert V
\rangle_{s3}+ \beta \vert V \rangle_{r3} \vert H
\rangle_{s4})\nonumber\\&&+ \eta_1 \eta_2(\alpha \vert H
\rangle_{r3} \vert V \rangle_{s4}+ \beta \vert V \rangle_{r4} \vert
H \rangle_{s3})].
\end{eqnarray}
One can see the last eight terms are preserved against the noise and
all of them have the same form $\alpha \vert H \rangle_{ri} \vert V
\rangle_{sj}+ \beta \vert V \rangle_{rj} \vert H \rangle_{si}$
($i,j\in\{3,4\}$) excepts the spatial modes. Bob's decoder, which is
used to eliminate the difference between these two photons and
extract the initial quantum state, is made up of frequency beam
splitter (FBS), frequency shifter (FS) and PBS. The FBS makes the
photons with the frequency $\omega_r$ and $\omega_s$ go through two
different paths directly \cite{fbs}. This operation can also be
implemented by wavelength-division-demultiplexing (WDM) \cite{fre}
or fiber bragg grating (FBG) \cite{freqkd3,freqkd4}.

After the operation of HWP placed on the up path, the frequencies of
these two photons will be manipulated to a same value set in advance
by means of the frequency shifter (FS). The modulation of the
frequency can be carried out by acousto-optic modulator (AOM)
\cite{fbs,aom}, sum-frequency generation (SFG) process \cite{sfg},
and so on. Then the distinguishability of these two photons are
eliminated and the interference could take place at the PBS$_3$. The
ultimate form of the total state is
\begin{eqnarray}
\vert \psi_3\rangle &=& \frac{1}{\sqrt{2}}[\alpha \delta_1^2\vert H
\rangle_{3y}\vert V \rangle_{3y} + \beta \eta_2^2\vert V
\rangle_{3x}\vert H \rangle_{3x} \nonumber\\&&+ \alpha \eta_1^2\vert
H \rangle_{4y}\vert V \rangle_{4y} + \beta
\delta_2^2\vert V \rangle_{4x}\vert H \rangle_{4x} \nonumber\\
&&+ \alpha \delta_1\eta_1(\vert H \rangle_{3y}\vert V \rangle_{4y} +
\vert H \rangle_{4y}\vert V \rangle_{3y})\nonumber\\&&+
\beta\delta_2 \eta_2 (\vert V \rangle_{3x}\vert H \rangle_{4x}
+ \vert V \rangle_{4x}\vert H \rangle_{3x})\nonumber\\
&&+ \delta_1 \eta_2(\alpha \vert V \rangle_{3x} \vert V
\rangle_{3y}+ \beta \vert H \rangle_{3y} \vert H
\rangle_{3x})\nonumber\\&&+ \eta_1 \delta_2(\alpha \vert V
\rangle_{4x} \vert V \rangle_{4y} + \beta
\vert H \rangle_{4y} \vert H \rangle_{4x})\nonumber\\
&&+ \delta_1 \delta_2 (\alpha \vert V \rangle_{4x} \vert V
\rangle_{3y}+ \beta \vert H \rangle_{3y} \vert H
\rangle_{4x})\nonumber\\&&+ \eta_1 \eta_2(\alpha \vert V
\rangle_{3x} \vert V \rangle_{4y}+ \beta \vert H\rangle_{4y} \vert H
\rangle_{3x})]. \label{psi4}
\end{eqnarray}
Here $x$ and $y$ are the two spatial modes of each PBS on the paths
3 and 4. We omit the subscript $r$ and $s$ as the frequency
distinguishability of these two photons is erased and they are
indistinguishable when they arrive at PBS$_3$ and PBS$_4$. From
Eq.(\ref{psi4}), we can select the instances that two photons
arriving at different spacial mode $x$ and $y$ ($3x/3y$, $4x/4y$,
$3x/4y$ and $4x/3y$) and measure one photon with $X$ basis, and then
the other photon can be manipulated to the initial state with a
proper unitary operation chosen according to the measurement result.
For example, if the outcome of the measurement on the photon coming
from $3y$ is $\vert +x\rangle_{3y}=\frac{1}{\sqrt{2}}(\vert H\rangle
+\vert V\rangle)$, the photon in the output $3x$ is in the state
$\alpha \vert V\rangle + \beta \vert H\rangle$ and Bob can obtain
the original state $\alpha \vert H\rangle + \beta \vert V\rangle$
with a bit-flipping operation $\sigma_x=|H\rangle\langle V| +
|V\rangle\langle H|$; If Bob obtains the outcome $\vert
-x\rangle_{3y}=\frac{1}{\sqrt{2}}(\vert H\rangle -\vert V\rangle)$,
the photon in the output $3x$ is in the state $\alpha \vert V\rangle
- \beta \vert H\rangle$ and Bob can obtain the original state
$\alpha \vert H\rangle + \beta \vert V\rangle$ with the operation
$-i\sigma_y=|H\rangle\langle V| - |V\rangle\langle H|$.

The success probability of this transmission scheme is $\eta^2/2$
with the help of the frequency DOF, where $\eta$ is the efficiency
of a FS. We also can simplify this protocol by using only one FS for
one decoder. That is keeping the frequency of the signal qubit and
adjusting the frequency of reference one to $\omega_s$. In this way,
the success probability is $\eta/2$, which can be 1/2 at best with
the efficient operation of FS.

\section{discussion and summary}

The main part of this qubit transmission scheme is the decode
process, in which two photons with different frequencies should be
guided to two different pathes faithfully and be reset to have the
same frequency by FS. The FS can be implemented by several means
with current technique, such as acousto-optic modulator (AOM
)\cite{fbs,aom}, sum-frequency generation (SFG) process\cite{sfg},
and so on. And sufficient high modulation efficiency is accessible.
It is reported in Ref. \cite{efficiency}, the internal conversion
efficiency of a SFG process is 99\% and the overall efficiency can
be 65\%. Moreover, the frequency difference between the signal
photon and the reference one should be set to a small value which
could be discriminate by the FBS effectively  and  the noise effects
on these two photons with different frequencies can be viewed as
identical.

Compared with PRL95 protocol, we use the frequency DOF to mark these
two photons instead of temporal DOF. The success probability of our
protocol can be eight times of PRL95 scheme at best, whose success
probability is only 1/16 without resorting to deterministic
two-qubit operation. The improvement comes from two parts. First, a
HWP is introduced in path 3 before the decoder, which is utilized to
extract the original state from terms that each outport of PBS$_2$
has one photon\cite{prl95}. With the use of HWP$_0$ shown in Fig.1,
the success probability of PRL95 scheme can be improved from 1/16 to
1/8 without any deterministic two-qubit operation. Second, the use
of frequency DOF avoids the 75\% loss during the eliminating of the
distinguishablity of these two photons.

Compared with the quantum state transmission scheme with single
photons \cite{apl91}, our protocols have no predominance in terms of
success probability. However, schemes with other DOF requires a much
smaller order of timing precision. Moreover, as the two photons in
cases we selected (x/y) are always in the same pure entangled state
$\alpha \vert VV \rangle + \beta \vert HH \rangle$, the scheme using
additional qubit can also be used to prepare entangled states
remotely in Bob's lab without classical communication with the
collective noise. This protocol can be used  not only  to transmit a
pure single-particle state, but also to transmit a mixture state.

In the quantum state transmission scheme with an additional qubit,
an assistant DOF, which is stable during the transmission, is needed
to differentiate the two photons for the sake of performing
different operations on each photon after the noise. Subsequently,
the DOF should be erased efficiently for extracting the uncorrupted
polarization state we want. The frequency DOF satisfies these
restrictions in principle. With the development of experiment
technique, more suitable DOF may be found in future and the
implementation of this theoretic protocol would be more efficient.

In summary, we have proposed an efficient faithful qubit
transmission scheme against collective noise with frequency DOF. The
success probability  can be 1/2 in principle, which is eight times
of success probability of PRL95 scheme utilizing the temporal DOF.
We also find a HWP is enough to improve success probability of PRL95
from 1/16 to 1/8 without using deterministic two-qubit operation.
The protocol we present is a theoretic frame and the success
probability of 1/2 can be obtained with the development of
experiment techniques. Maybe some other DOFs which are antinoise and
easy to control will be found in the future. A suitable DOF for
assisting the transmission of polarization state can be chosen
according to the practical condition.

\section*{ACKNOWLEDGEMENTS}
This work is supported by the National Natural Science Foundation
of China under Grant No. 10604008, a Foundation for the Author of
National Excellent Doctoral Dissertation of China under Grant No.
200723, and Beijing Natural Science Foundation under Grant No.
1082008.

\end{document}